\documentclass[showkeys,amsmath, amssymb, aps, final,floatfix]{revtex4}

\usepackage{graphics}
\newcommand{\mb}[1]{\mathbf{#1}}

\newcommand{\df}{\displaystyle\frac}
\newcommand{\rmd}{\mathrm{d}}
\newcommand{\etal}{{et al.}}

\begin{document}

\title{Unwinding of a cholesteric liquid crystal and bidirectional surface anchoring} 
\author{G. \surname{McKay}\email{g.mckay@strath.ac.uk}}  
\affiliation{Department of Mathematics and Statistics, University of Strathclyde,
 University of Strathclyde,
 Glasgow G1~1XH, U.K.}


\begin{abstract}
\noindent We examine the influence of  bidirectional anchoring  on the unwinding of a planar cholesteric liquid crystal induced by  the application of a magnetic field. We consider a  liquid crystal layer confined between two plates with the helical axis perpendicular to the substrates. We fixed the director twist on one boundary and   allow for bidirectional anchoring on the other by introducing a  high-order surface potential. 
By minimizing the total free energy for the system, we investigate the untwisting of the cholesteric helix as the liquid crystal attempts to align with the magnetic field. The transitions between metastable states occur as a series of pitchjumps as the  helix expels quarter or half-turn twists, depending on the relative sizes of the strength of the  surface potential and the bidirectional anchoring.
We show  that secondary easy axis directions  can play a significant role in the unwinding of the cholesteric in its transition towards a nematic, especially when the surface anchoring 
strength is large. 
\end{abstract}
\maketitle



\section{Introduction}

A cholesteric (or chiral nematic) is a type of liquid crystal whose  chiral nature     causes the constituent molecules to align at a slight angle to one another. This leads to a periodic configuration in which the preferred direction of the long molecular axis (or  director)  twists continuously in space perpendicular to a helical axis. The length over which the director rotates by $2\pi$ radians is known as its pitch and can vary from $200\,$nm  upwards~\cite[p.5]{deJEU}.  In the absence of any external influences such as an applied field,  the cholesteric  possesses a natural or equilibrium pitch that depends on the temperature of the liquid crystal. However, due to diamagnetic or dielectric anisotropy, the period of the helical structure can be changed by the application of a magnetic or  electric field. 
 De Gennes~\cite{deGENNES} and Kedney and Stewart~\cite{KEDNEY}  predict theoretically how the  helix can be completely unwound in an infinite sample of cholesteric liquid crystal, resulting in a cholesteric to homeotropic (planar) nematic phase transition.
 The same field-induced transition is also observed experimentally by Meyer~\cite{MEYER}. Subsequent studies  examine the dependence of  the observed helical pitch on the field strength and critical fields for complete unwinding~\cite{WYSOCKI,MEYER2,DURAND,BAESSLER}, and  allow experimentalists to measure physical quantities, for example, the twist elastic constant of the cholesteric. 
 
When considering an infinite sample of cholesteric (i.e.~a relatively thick sample in which the bulk is unaffected by any boundary surfaces), the pitch changes continuously, increasing smoothly with the applied field until it becomes infinite and the helix is completely unwound. However, when the liquid crystal has a finite thickness, confined between two substrates with some degree of anchoring on the surfaces, changes in pitch may occur in discrete jumps. These pitchjumps can arise due to changes in the natural pitch with temperature~\cite{Pink,Gandhi,AndBau} and are often associated with thermal hysteresis~\cite{PAL,Yoon,McKay}. An applied field can also lead to stepwise changes in pitch and helix unwinding in confined  samples~\cite{DREHER,vanSPRANG3,Schlangen,Smalyukh}. Kedney and Stewart~\cite{KEDNEY2}
present a theoretical analysis of the unwinding of a cholesteric with strong anchoring on the substrates, i.e.~the angle of director twist is fixed on the boundaries. As we will discuss in Sect.~\ref{RESULTS}, different metastable states can coexist for a given field strength. The discrete pitchjumps  coincide with a change in nature of the twist profile that provides the global energy minimizer. More recently, Scarfone~\etal~\cite{SCARFONE} generalize the problem of \cite{KEDNEY2} to consider an in-plane magnetic field tilted at some angle with respect to  fixed parallel twist directions on the substrates. The analysis of Lelidis~\etal~\cite{LELIDIS}  allows for  an incomplete number of half twists in the liquid crystal layer by imposing strong homogeneous anchoring with  non-parallel director  twists on the two confining plates. 

More realistic boundary conditions for liquid crystals allow for the  director angle on a boundary to vary because of the competition between the bulk alignment
and a preferred surface direction (or easy axis direction). The director is thought to be weakly anchored at the substrate, with a degree of flexibility controlled by a finite anchoring strength combined with a surface energy. 
Easy axis directions can be imposed on solid substrates via a variety of methods, for example, surface rubbing and oblique evaporation of a SiO thin film on the surface.
As the anchoring strength increases, we revert to  strong boundary conditions  with the direction fixed in the easy axis direction on the substrate, also known as infinite anchoring.
Belyakov and co-workers present theoretical analyses of the untwisting of a cholesteric due to the action of a field or temperature~\cite{BEL1,ZB1,ZB2,ZB3,BSO1,BSO2} with weak anchoring on the bounding plates, whereas the stability of the helical structures when there is   asymmetry due to different anchoring strengths on the two surfaces is considered by Kiselev and Sluckin~\cite{KIS}. 

Most of the studies examining  discrete jumps in  pitch in cholesteric liquid crystal cells bounded by two parallel substrates employ an anchoring potential of the form
\begin{equation}
w_\mathrm{s}(\bar{\phi})= \df{1}{2}\tau_0\sin^2\bar{\phi} \label{rp1}
\end{equation}
on one or both plates,
where $\tau_0$ is the anchoring strength and $\bar{\phi}$ is the director azimuthal twist angle at the surface. This is the twist equivalent of the quadratic  surface energy density first proposed by Rapini and Papoular~\cite{RAPINI} and adopted widely in models for liquid crystals~\cite[p.49]{Stewart}. The form (\ref{rp1})  represents a 
substrate that is rubbed  to provide  easy axes for the director at $\bar{\phi}=k\pi$ radians, where $k$ is an integer. The quadratic expansion also ensures that the 
 inversion symmetry of cholesterics is preserved. It is also possible, however,  to obtain bidirectional surface ordering in liquid crystal devices with two easy directions on a substrate. This  
 can be achieved via a variety of treatments at the upper surface plate, for example,  patterned surfaces \cite{Yoneya1,Fukuda1,Fukuda2}, SiO evaporation~\cite{Sergan} and non-parallel aligning films~\cite{Barberi}.
Mathematically, bistable surface anchoring can be modelled by introducing a higher order surface potential  into (\ref{rp1}).  The  theoretical studies of Sergan and Durand~\cite{Sergan}, Barberi~\etal~\cite{Barberi} and Yoneya~\etal~\cite{Yoneya2} incorporate a quartic expansion in $\sin\bar\phi$,
\begin{equation}
w_{\mathrm{s}}(\bar{\phi})= \df{1}{2}\tau_0( \sin^2\bar{\phi} + \zeta\sin^4\bar{\phi}).\label{WSEN}
\end{equation} 
The dimensionless bidirection coefficient $\zeta$  depends on the nature of the interaction between the liquid crystal and the surface, with   $\zeta=0$ corresponding to the  quadratic  Rapini-Papoular anchoring~(\ref{rp1}).
The higher-order potential (\ref{WSEN}) still preserves the inversion symmetry of the cholesteric but also provides secondary easy directions corresponding to odd multiples of $\pi/2$ radians when  $\zeta<-1/2$. In particular,  $\zeta=-1$ provides surface potential minima of equal strength at all integer multiples of $\pi/2$ radians.
The quartic form (\ref{WSEN}) has been generalized by Pieranski and J\'er\^ome~\cite{Pieranski} in a study of discontinuous first-order anchoring transitions by introducing a phase angle in the fourth-order term.
McKay~\cite{McKay} employs (\ref{WSEN}) in a study of the thermal hysteresis of pitchjumps in a planar cholesteric and discusses how  the high-order term can still alter the pitchjump process even when $\zeta>-1/2$. 
Apart from an initial discussion about the Rapini-Papoular case $\zeta=0$, 
here we concentrate on perpendicular easy directions and  $\zeta\approx -1$.

The aim of this paper is to examine the influence of a bidirectional anchoring potential on the unwinding of a cholesteric liquid crystal subject to the application of a magnetic field. We adopt the quartic surface energy (\ref{WSEN}) on the upper boundary confining a layer of cholesteric, while maintaining a strong anchoring condition on the lower surface. In Sect.~\ref{model} we introduce the model for the liquid crystal layer, including the elastic energy density and total energy per unit area.  We then derive the differential equations from which we  obtain the director twist across the layer. Section~\ref{RESULTS} examines the unwinding of the cholesteric through a series of pitchjumps at critical values of the field strength. These can be quarter or half-turn changes in the twist angle depending on the choice of bidirectional anchoring parameter $\zeta$ and the anchoring strength. We show that the influence of the bidirectional anchoring increases as the anchoring strength increases, although it may still be possible for the unwinding to bypass intermediate easy axis directions when the magnitude of the field is relatively large and the cholesteric is almost completely unwound.

\section{Model}\label{model}

\begin{figure}
\centerline{\resizebox{9cm}{!}{\includegraphics{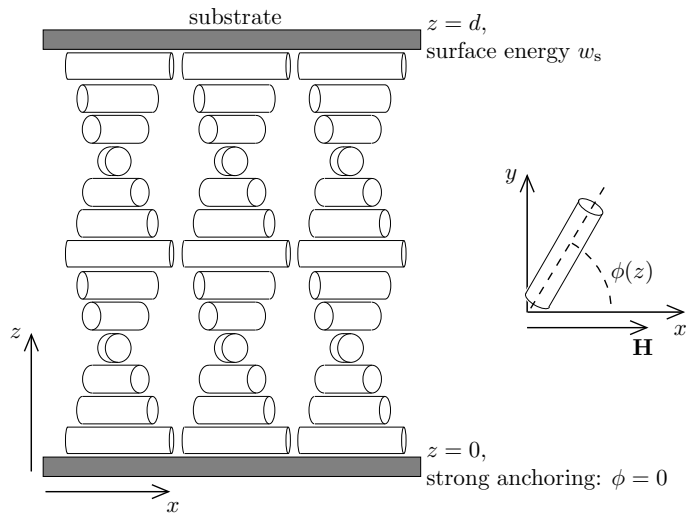}}}
 \caption[]{Cholesteric liquid crystal confined between two plates at $z=0$ and $z=d$ with its helical axis in the $z$-direction. The director twist with respect to the $x$-axis direction is $\phi(z)$, while $\mb{H}$ is the in-plane  magnetic  field. 
 }\label{fig1} 
 \end{figure}

\noindent We consider a cholesteric liquid crystal of thickness $d$ between two boundary plates at $z=0$ and $z=d$. Assuming that the nematic director lies in the $xy$-plane and the helical axis is in the $z$-direction, the director can be described via
\begin{equation}
\mb{n}=\bigl(\cos\phi(z),\,\sin\phi(z),\,0\bigr),\label{DIR}
\end{equation}
where $\phi(z)$ is the director twist angle measured with respect to the $x$-axis,
 as shown in Figure~\ref{fig1}. 
The liquid crystal is subject to an in-plane magnetic field $\mb{H}=H(1,\,0,\,0)$ of 
magnitude $H(\ge 0)$.
Combining the magnetic and elastic energy densities, we can express the overall bulk energy density for the cholesteric~\cite[chapter 6]{deGPro} as
\begin{eqnarray}
w_\mathrm{b}
&=& \df{1}{2}K_2 \Big( \mb{n}\cdot\nabla\times\mb{n}   - \df{2\pi}{p}\Big)^2 -\df{1}{2} \chi_a ( \mb{n}\cdot\mb{H})^2 \nonumber \\
&=&\df{1}{2}K_2 \Big( \df{\rmd \phi}{\rmd z} - q \Big)^2 -\df{1}{2}\chi_a H^2\cos^2(\phi
),\label{WBEN}\end{eqnarray}
where $K_2$ is the elastic constant associated with twist of the cholesteric and  $\chi_a$ is the magnetic anisotropy, here assumed to be positive so that the liquid crystal director prefers to align with the field. The wavenumber  $q = {2\pi}/{p}$ is also assumed to be positive so that 
 the cholesteric exhibits a right-handed helix.  The natural, or equilibrium, pitch $p$ is the distance along the helical axis over which the director twists $2\pi$ radians in the absence of the  applied field or surface anchoring.   
On the lower plate at $z=0$ we assume that the director is fixed such that $\phi(0)=0$. At the upper surface we introduce the  bidirectional surface energy (\ref{WSEN}), 
where $\overline\phi$ represents the twist on the substrate.

Combining the bulk and surface energies, we can now construct the total energy of our system per unit area,
\begin{equation}
W= \int_0^d w_\mathrm{b}\; \mathrm{d}z + w_\mathrm{s},\label{DIMENERGY}
\end{equation}
where $w_\mathrm{s}$ is the quartic surface potential (\ref{WSEN}).
Equilibrium profiles for the director twist can be found by minimizing the total energy $W$ with respect to the angle $\phi$. Before doing this, we first non-dimensionalize (\ref{DIMENERGY}) by rescaling $z \to z/d$ and introducing a  modified total energy 
\begin{eqnarray}
\widehat{W}=\df{2d}{K_2}W  
&= &\int_0^1 \Bigl( \df{\mathrm{d} \phi  }{\mathrm{d} z } - \pi \hat{q}\Bigr)^2 -\lambda^2 \cos^2(\phi  
)\; \mathrm{d}z  + \df{2\pi}{\rho}\hat{w}_\mathrm{s}\nonumber\\
&\equiv&  \int_0^1     \hat{w}_\mathrm{b}    \; \mathrm{d}z      + \df{2\pi}{\rho}\hat{w}_\mathrm{s},
\label{EN2}
\end{eqnarray}
where the dimensionless surface energy $\hat{w}_\mathrm{s} =\bigl( \sin^2\overline{\phi} + \zeta\sin^4\overline{\phi}\bigr)/2$ and $\bar{\phi}$ still represents the director twist on the upper surface.  We have also introduced non-dimensional parameters
$$\rho=\df{\pi K_2}{d\tau_0},\qquad \hat{q}=\df{q d}{\pi},\qquad  \lambda^2=\df{d^2\chi_a H^2}{K_2}.$$
Non-negative  $\rho$ is a rescaled reciprocal of the anchoring strength, with $\rho=0$ corresponding to infinite anchoring. 
The parameter $\hat{q}$ represents the   number of half (or $\pi$)-twists 
in a sample of depth $d$  if the director was allowed to rotate freely on the upper plate (zero anchoring) and  the magnetic field is switched off. The helix will attempt to unwind as the magnetic field strength increases,  so the actual number of half-twists exhibited by the cholesteric may differ from $\hat{q}$. In order to focus on the competition between the field strength and surface anchoring, we have fixed $\hat{q}=10$ in Figures~\ref{figX1} to \ref{fig6}.
Finally, $\lambda$ is a measure of the magnitude of the magnetic field  relative to the twist elastic constant.

We  calculate the equilibrium twist profiles for the cholesteric by minimizing the energy $\widehat{W}$.  
The Euler-Lagrange equation derived from (\ref{EN2}) is 
\begin{equation}
\df{\mathrm{d}^2 \phi}{\mathrm{d}z^2} -\lambda^2 \cos(\phi
)\sin(\phi
)=0,\qquad z\in(0,\,1).\label{EQC}
\end{equation}
The boundary condition for the twist at the upper plate can also be obtained from calculus of variations~\cite[chapter 4]{Courant},
\begin{equation}\frac{\partial  \hat{w}_\mathrm{b}}{\partial \phi'}+ \frac{\mathrm{d} \hat{w}_\mathrm{s}}{\mathrm{d}
\overline\phi}=0\  \mbox{ on }\ z=1, \label{BCW}\end{equation}
where   $\phi'=\mathrm{d}\phi/\mathrm{d}z$. Substituting  $\hat{w}_{\mathrm{b}}$ and $\hat{w}_{\mathrm{s}}$ defined in (\ref{EN2}) into (\ref{BCW}), we can write the boundary condition for $\phi$ incorporating weak anchoring as
\begin{equation}\df{\mathrm{d} \phi}{\mathrm{d}z} - \pi \hat{q} + \df{\pi}{\rho}
\sin\overline\phi \cos\overline\phi (1+2\zeta\sin^2\overline\phi)=0
\   \mbox{ on }\  z=1. \label{EQB} \end{equation}
Equilibrium twist profiles are now solutions of (\ref{EQC}) and (\ref{EQB}), in conjunction with the condition that the angle vanishes at $z=0$. 

Following a procedure similar to that adopted in Kedney and Stewart~\cite{KEDNEY} and Scarfone~\etal~\cite{SCARFONE},
we can obtain an implicit form for $\phi(z)$ from (\ref{EQC}),
%
%
\begin{equation}
 {\cal{F}}(\phi,\, k
 )-\lambda z   =0,\label{EQG}
\end{equation}
where $k(>0)$ is a constant of integration to be determined,
\begin{equation}
{\cal{F}}(\phi,\, k
)=\int_0^\phi   \df{\mathrm{d}\psi}{ \sqrt{ k+ \sin^2
\psi
}}=\frac{1}{\sqrt{k}} F\big (\phi\,|\, -k^{-1}\big)\nonumber
\end{equation}
and $F(\phi\,|\,m)$ is   the incomplete elliptic integral of the first kind.  
The limiting value of $z=1$ in (\ref{EQG})  provides an implicit form relating $k$ and  $\overline\phi=\phi(1)$, the twist on the upper plate,
\begin{equation}{\cal{F}}(\overline\phi,\, k
)-\lambda=0.\label{EQE}
\end{equation}
However, if we replace the derivative at $z=1$ in (\ref{EQB}) by the term derived from (\ref{EQG}), we can  obtain another expression for the constant $k$ in terms of $\overline\phi$, namely 
\begin{equation}
k= \df{1}{\lambda^2} \Bigl(\pi \hat q - \df{\pi}{\rho} \df{\mathrm{d} \hat{w}_\mathrm{s}}{\mathrm{d} \overline\phi}\Bigr)^2- \sin^2
\overline\phi 
.\label{EQF}
\end{equation}
Together, (\ref{EQE}) and (\ref{EQF}) provide the constant $k$ and the angle $\overline{\phi}$ corresponding to the chosen non-dimensional parameters $\rho$, $\hat{q}$ and $\lambda$. These lead, in turn, to an implicit form for $\phi(z)$ from (\ref{EQG}).

\section{Discussion}\label{RESULTS}

\begin{figure}
\centerline{\resizebox{11cm}{!}{\includegraphics{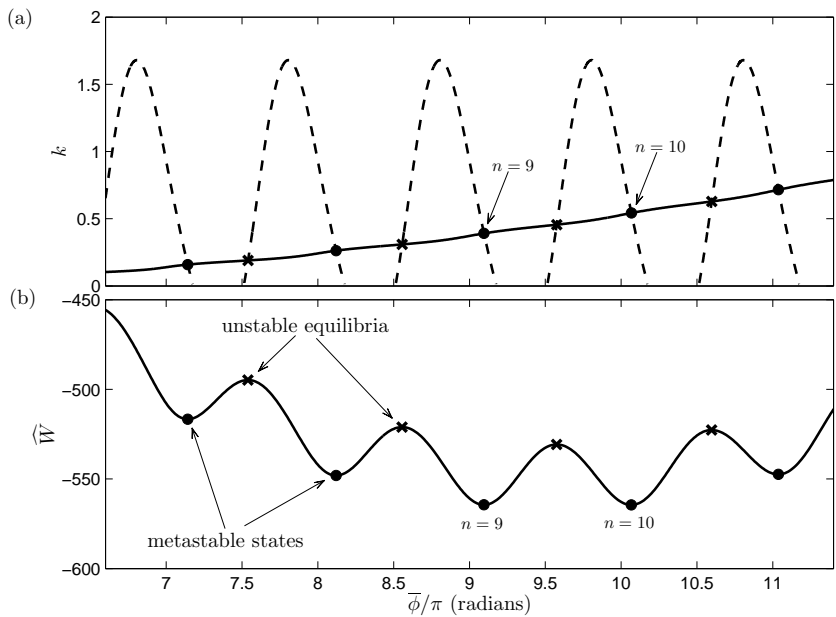}}}
 \caption[]{(a) Constant of integration $k$ derived from implicit equation (\ref{EQE}) (solid line) and boundary condition (\ref{EQF}) (dashed line) as angle $\overline\phi$ varies, with $\lambda=32.58$, $\rho=0.1$, $\zeta=0$ and $\hat q=10$;   (b) Energy $\widehat W$, where $k$ is calculated using the implicit form (\ref{EQE}). The intersections of the $k$-curves coincide with energy extrema and allow us to calculate  the energy profile  $\phi(z)$ via (\ref{EQG}).  }\label{figX1}  
 \end{figure}

\begin{figure}
\vspace*{.1cm}
\centerline{\resizebox{10cm}{!}{\includegraphics{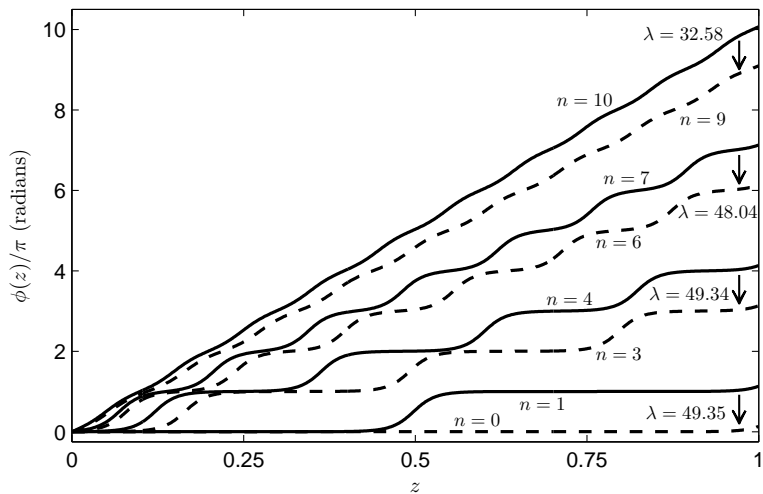}}}
 \caption[]{Unwinding of the director twist $\phi(z)$ at specific values of parameter $\lambda$ with $\rho=0.1$, $\zeta=0$ and $\hat q=10$. 
 In each case, the chosen $\lambda$ corresponds to a critical value where the helix expels a half-twist as the cholesteric unwinds. For each profile, $n$ represents  the number half-twists rounded to the nearest integer multiple of 0.5.}\label{fig2}  
 \end{figure}

\begin{figure}
\centerline{\resizebox{10cm}{!}{\includegraphics{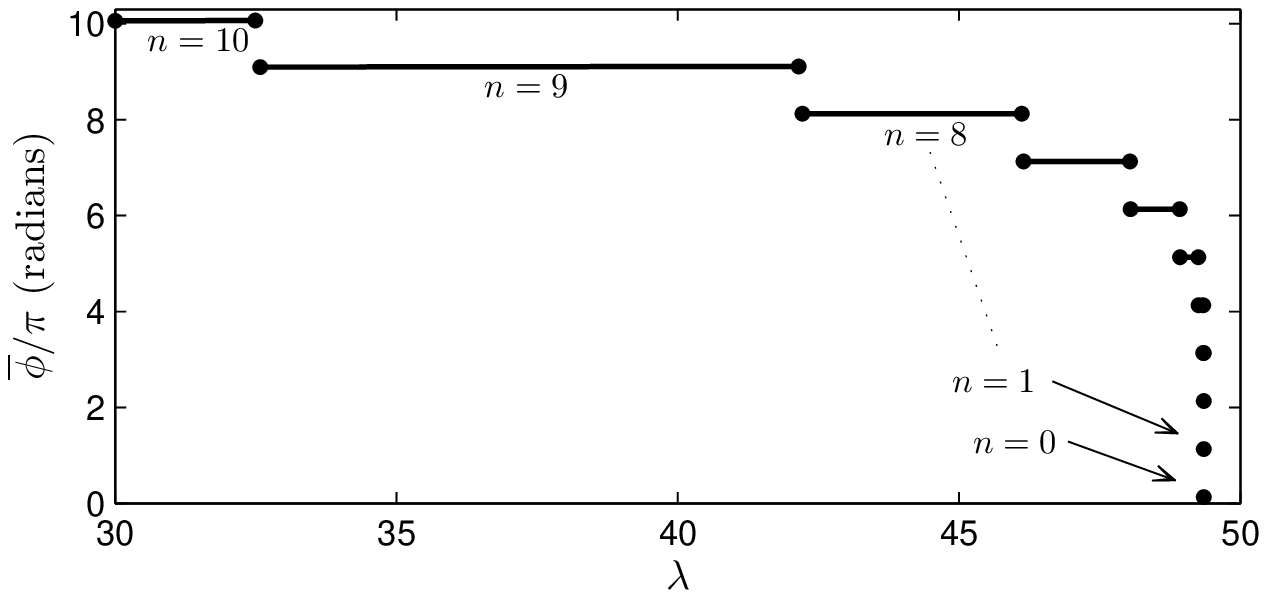}}}
 \caption[]{Azimuthal twist angles at the upper plate for variable parameter $\lambda$, with  $\rho=0.1$, $\zeta=0$ and $\hat q=10$. 
 As the field strength increases, the global minimum energy state expels (approximate) half-twists and the cholesteric helix unwinds. The cascade of pitchjumps from $n=4$ to $n=0$ takes place over a very short interval  when $\lambda$ is large and the helix is nearly unwound. Note, as seen in Figure~\ref{fig2}, the twist angles at the upper surface are close but not equal to integer multiples of $\pi$ radians because of the weak anchoring condition. }\label{fig3}  
 \end{figure}

\noindent In Figure~\ref{figX1}(a) we plot the constant $k$ obtained  from both the implicit form    (\ref{EQE}) and condition (\ref{EQF}) for the simple Rapini-Papoular energy (\ref{rp1})  as $\overline\phi$ varies. Note that the graph derived from (\ref{EQF}) is restricted to the values of $\overline\phi$ that provide $k>0$. Figure~\ref{figX1}(b) represents the non-dimensional energy $\widehat{W}$ introduced in (\ref{EN2}) for the same range of $\overline\phi$, with the values of $k$ calculated using (\ref{EQE}). The intersections of the curves in Figure~\ref{figX1}(a) correspond to energy extrema in Figure~\ref{figX1}(b).
The magnetic field contribution to the total energy  is minimized when the director is aligned in directions which are integer multiples of $\pi$ radians. 
Therefore, as the magnetic field strength increases and dominates elastic or weak anchoring effects, 
 the  cholesteric undergoes a series of transitions as its helix unwinds in an attempt to align with the magnetic field. Since multiple metastable states can coexist for a given $\lambda$,  the pitchjumps coincide with 
 discrete changes in the overall twist  of the  energy global  minimizer.
 The value of $\lambda$ chosen in Figure~\ref{figX1} coincides with a pitchjump as the helix expels approximately a half (or $\pi$)-twist and aligns with the field in more of the cell. 
Figure~\ref{fig2} plots the equilibrium twist profiles $\phi(z)$ obtained from (\ref{EQG})--(\ref{EQF}) for a sequence of critical values of the parameter $\lambda$.  
In the analysis that follows we refer to \emph{quarter} and \emph{half-turn} changes in the overall twist across the entire cell, i.e.~variations in the director angle at the upper plate. In reality, these jumps will not be exact integer multiples of $\pi/2$ or $\pi$ radians, respectively, because of the weak anchoring.   We can categorize each profile in Figures~\ref{figX1} and \ref{fig2} by $n$,  the number half-twists it possesses, or simply $\overline\phi/\pi$, rounded to the nearest integer multiple of 0.5. For example, the pitchjump at $\lambda=32.58$ corresponds to a transition from a $n=10$ to a $n=9$ state. 
 Figure~\ref{fig3} shows the full cascade of transitions  as the field strength increases until the cholesteric is virtually completely unwound, although
  a small residual twist remains at the upper surface  for the $n=0$ state due to the finite surface energy and elastic effects. As the magnetic field strength is increased even further, this residual surface twist will decrease towards zero.

\begin{figure}
\centerline{\resizebox{11cm}{!}{\includegraphics{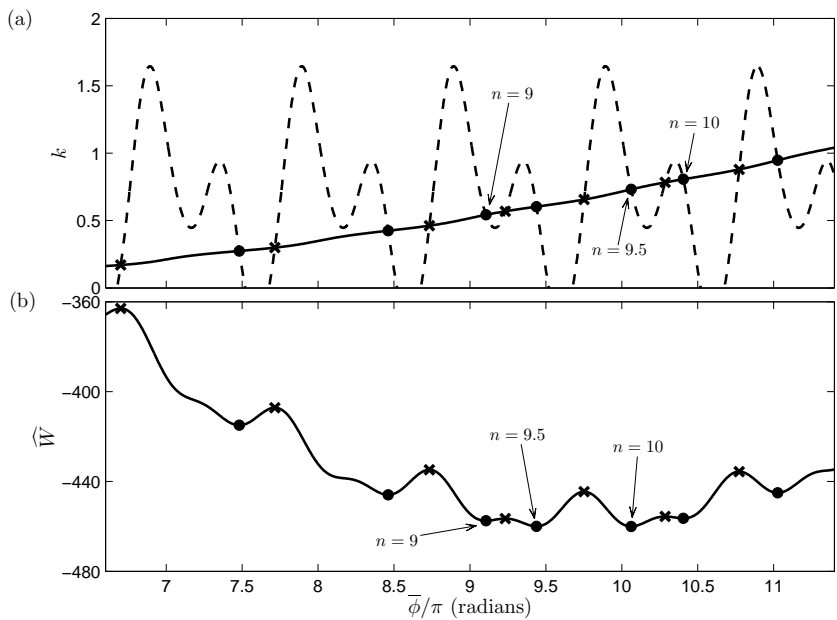}}}
 \caption[]{ (a) Constant of integration $k$ derived from implicit equation (\ref{EQE}) (solid line) and boundary condition (\ref{EQF}) (dashed line) as angle $\overline\phi$ varies, with $\lambda=29.50$, $\rho=0.1$, $\zeta=-1$ and $\hat q=10$.  (b) Energy $\widehat W$, where $k$ is calculated using the implicit form (\ref{EQE}). Oscillations in the $k$-curve obtained from  (\ref{EQF}) result in  secondary metastable states.  }\label{figX2}  
 \end{figure}

\begin{figure}

\centerline{\resizebox{10cm}{!}{\includegraphics{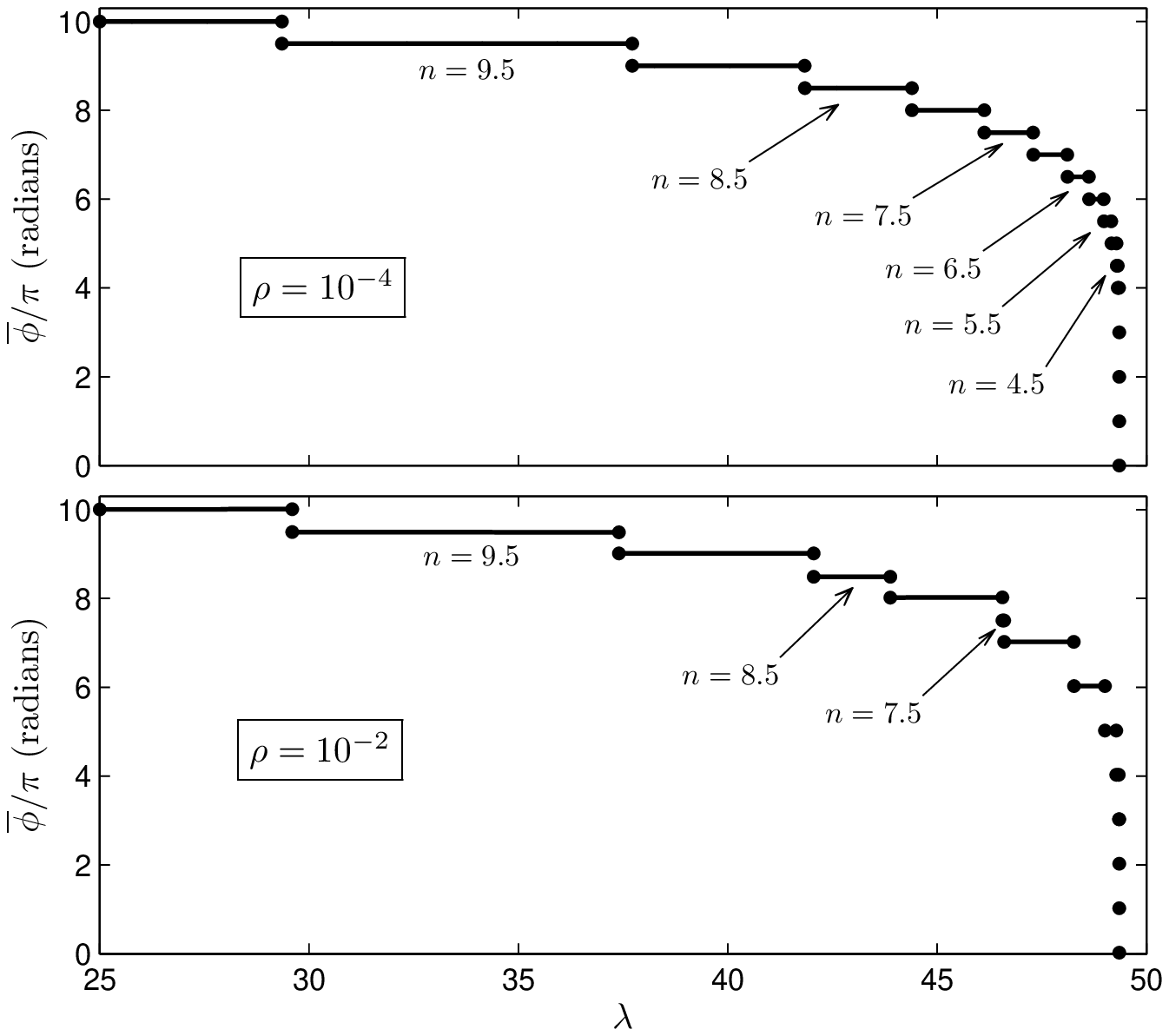}}}
 \caption[]{Azimuthal twist angles at the upper plate for $\zeta=-1$, $\hat{q}=10$ and  variable  $\lambda$. The surface twist now displays values close to the secondary easy directions at odd multiples of $\pi/2$ radians as the helix untwists.
 When $\rho=10^{-4}$ the secondary pitchjumps persist until
  the final transitions when $\lambda$ is large.  For $\rho=10^{-2}$, which corresponds to  weaker anchoring, only the higher-twist intermediate states are observed as the helix unwinds.    }\label{fig4}  
 \end{figure}

The quadratic term in the surface energy (\ref{WSEN}) is minimized when the surface twist aligns at an integer multiple of $\pi$ radians, in a fashion similar  to the director in the bulk of the liquid crystal cell when acted upon by the field.
However, if we introduce bidirectional surface anchoring by including the quartic term in (\ref{WSEN}) for $\zeta<-1/2$, then the new intermediate surface energy minima at odd multiples of $\pi/2$ radians will compete with the magnetic field alignment. 
Figure~\ref{figX2} shows the intersections of the $k$-curves and the corresponding energy $\widehat W$ in the presence of bidirectional anchoring with $\zeta=-1$. The oscillations in the curve obtained from (\ref{EQF}) result in intermediate twist  profiles and secondary metastable states corresponding to $n=10.5,\,9.5,$ etc. 
This is illustrated further in Figure~\ref{fig4}    for contrasting values of the  surface anchoring parameter $\rho$.  
For the relatively strong anchoring condition ($\rho=10^{-4}$), most of the  intermediate metastable states act as the global energy minimizer at some stage as  $\lambda$ increases.
The step unwinding of the cholesteric occurs in $\pi/2$ pitchjumps until the liquid crystal is almost fully unwound, with only the final  intermediate states $n=3.5$ to $n=0.5$ skipped when $\lambda$ is large.
Significantly,  for the weaker surface anchoring $\rho=10^{-2}$, a reduced number of the intermediate twists play a role in the cascade of pitchjumps. 
At higher magnetic fields, the cholesteric bypasses the secondary easy axis directions and unwinds in an extended series of  half instead of quarter-twist pitchjumps at the upper surface.


\begin{figure}
\centerline{\resizebox{\textwidth}{!}{\includegraphics{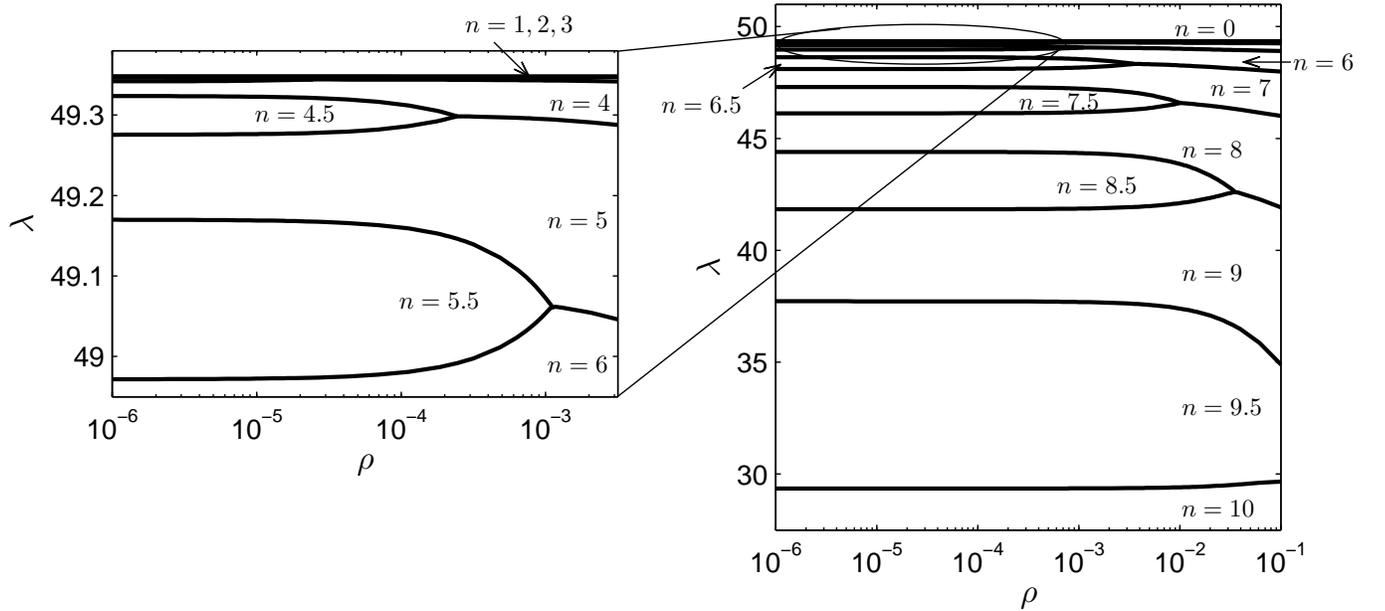}}}
 \caption[]{Regions in $(\lambda,\,\rho)$ space where the different $n$-states act as the global energy minimum for  $\zeta=-1$, $\hat{q}=10$. The states $n=1,\,2$ and $3$ occur    in the small region before the helix unwinds but cannot be distinguished in the figure.}\label{fig5}  
 \end{figure}

Figure~\ref{fig5}   examines the influence of the anchoring strength in determining whether the director twist will bypass one or more  of the intermediate metastable states as the helix unwinds. 
Critical values of $\lambda$ are plotted for each pitchjump transition and variable $\rho$. The branches of Figure~\ref{fig5} demarcate the regions in $(\lambda,\,\rho)$ space corresponding to the different $n$-states. For a fixed $\rho$, we can determine the sequence of unwound twists as $\lambda$ increases in a manner similar to Figures~\ref{fig3} and \ref{fig4}.  
We observe from  Figure~\ref{fig5}  that the intermediate states at odd multiples of $\pi/2$ radians play a diminishing role as $\rho$ increases.
When the anchoring strength is relatively weak, the magnetic and bulk elastic terms dominate the energy of the system, especially for large field strengths. 
The field prefers to align at integer multiples of $\pi$ radians and, 
consequently, the surface energy can no longer constrain the upper surface twist  to an angle close to a secondary easy axis direction. As a result, helix unwinding takes place via a series of half-twist pitchjumps focussed on the integer $n$ states. 
For mid-strength anchoring, quarter-turn pitchjumps may occur initially as the helix unwinds, but are bypassed at higher fields as also shown previously in Figure~\ref{fig4}.

\begin{figure}
\centerline{\resizebox{\textwidth}{!}{\includegraphics{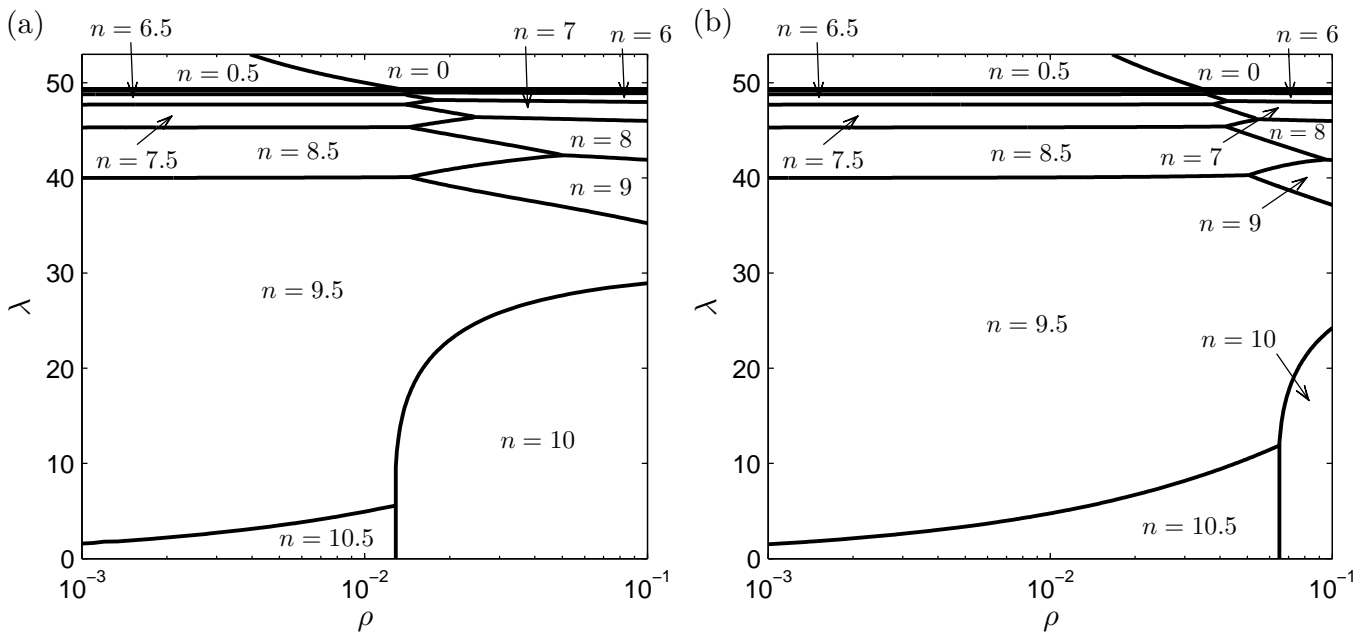}}}
 \caption[]{Regions in $(\lambda,\,\rho)$ space where the different $n$-states are the global energy minimum for $\hat{q}=10$:  (a) $\zeta=-1.01$;  (b) $\zeta=-1.1$. As $\zeta$ decreases the secondary states dominate the cholesteric transition to a planar nematic.
 }\label{fig6}  
 \end{figure}

The surface energy term in (\ref{EN2}) is very sensitive to the choice of $\zeta$, as can be seen by re-expressing $\hat{w}_{\mathrm{s}}$  in the form
\begin{equation}
  \hat{w}_\mathrm{s}=\df{1}{8}\sin2\bar\phi + \df{1}{2}(\zeta+1)\sin^4\bar\phi.\label{WX}
 \end{equation}
The first term in (\ref{WX}) vanishes for \emph{all} easy directions $\bar\phi=k\pi/2\ (k\in\mathbb{Z}).$ However, when combined with the $2\pi/\rho$ coefficient in (\ref{EN2}), the second term can be significantly large in magnitude for $\bar{\phi}$ close to the secondary easy axes  (odd integer $k$). Generally,  if $\zeta$ is  slightly greater than $-1$ then the surface energy contribution to (\ref{EN2})
is positive for all secondary states, and large enough to ensure that these secondary states never act as global energy minima, i.e.~the $n=9.5,\,8.5\,\ldots$ regions in Figure~\ref{fig5} shrink very rapidly as $\zeta$ increases from $-1$.
Conversely, the secondary states encroach further into $(\lambda,\,\rho)$ space even if  $\zeta$ is decreased by only a relatively small amount. 
In Figure~\ref{fig6} we consider the effect of a small decrease in the bidirection coefficient from $\zeta=-1$, biasing the surface energy towards the secondary directions at  odd multiples of $\pi/2$ radians on the boundary. Unlike the situation for $\zeta\ge -1$, this bias can lead to equilibrium states with $n=10.5$ when the magnetic field is relatively weak, and $n=0.5$ when $\lambda$ is large and the cholesteric is almost fully unwound.
More significantly, it is the secondary states which now play the dominant role and across a much wider range of $\rho$ than when $\zeta=-1$. Integer $n$ states that characterize the influence of the magnetic field are excluded from the  unwinding process until $\rho$ and $\lambda$ are relatively large.
%
Figures~\ref{fig5} and \ref{fig6} both illustrate that the secondary easy axis directions are prevalent when the anchoring strength is relatively large.  More significantly, the figures also show how even very small changes in the nature of the surface potential, with a shift in bias from primary to secondary easy axis directions, can affect the manner in which the helix unwinds.

\section{Conclusion}\label{Conclusion}
We have examined the unwinding of a planar cholesteric liquid crystal subject to bidirectional anchoring on its upper plate. By determining the states which minimize the total free energy described in terms of the director twist angle, we have modelled the unwinding of the cholesteric helix  via a series of near quarter or half-turn pitchjumps 
depending on the choice of bidirectional coefficient. In the transition to the nematic state, a competition exists between the twist angles favoured  by the magnetic field and  the easy axis directions imposed by the surface potential. Secondary easy axes can influence the unwinding when the surface anchoring strength is relatively strong and when the  potential is biased towards secondary twisted states via the coefficient $\zeta$. Although not considered here,
the behaviour of a cholesteric as it transitions to the nematic state could also be altered by a surface treatment which leads to non-perpendicular easy directions. 
 Another method of controlling the nature of the helix as it unwinds could be the application of an in-plane magnetic field that is tilted with respect to the easy axes, as considered by Scarfone~\etal~\cite{SCARFONE} for  Rapini-Papoular anchoring. For example, consider a magnetic field that is tilted at a specific angle and its strength increased until the cholesteric helix has unwound. If the tilt angle is then changed by  a small amount and the field strength decreased, the helix rewinding may be characteristically different from the unwinding process because the field is more closely aligned with a different easy direction.

\end{document}